# Enhanced low energy fluctuations and increasing out-of-plane coherence in vacancy ordered Na$_x$CoO$_2$

P. Lemmens,[2] V. Gnezdilov,[1,2] G. J. Shu,[3] L. Alff,[4] C. T. Lin,[5] B. Keimer,[5] F. C. Chou[3]

[1]*Inst. for Condensed Matter Physics, TU Braunschweig, D-38106 Braunschweig, Germany*

[2]*B.I. Verkin Inst. for Low Temperature Physics and Engineering, NASU, 61103 Kharkov, Ukraine*

[3]*Center for Condensed Matter Sciences, National Taiwan University, Taipei 10617, Taiwan*

[4]*Inst. of Materials Science, TU Darmstadt, Petersenstrasse 23, D-64287 Darmstadt, Germany*

[5]*Max-Planck-Institute for Solid State Research, Heisenbergstrasse 1, 70569 Stuttgart, Germany.*

We report Raman scattering experiments on the strongly correlated electron system Na$_x$CoO$_2$ with x= 0.71 and ordered Na vacancies. In this doping regime, Na$_x$CoO$_2$ exhibits a large and unusual thermopower and becomes superconducting upon hydration. Our Raman scattering data reveal pronounced low energy fluctuations that diverge in intensity at low temperatures. Related to these fluctuations is a drastic decrease of an electronic scattering rate $\Gamma$ from 50 to 3 cm$^{-1}$. This observation is quite different from the behavior of Na disordered samples that have a larger and temperature independent scattering rate. Simultaneously with the evolution of the scattering rate, phonon anomalies point to an increasing out-of-plane coherence of the lattice with decreasing temperature. These observations may indicate the condensation of spin polarons into an unusual, highly dynamic ground state.

PACS numbers:

71.30.+h Metal-insulator transitions and other electronic transitions
71.38.-k Polarons and electron-phonon interactions
78.30.-j: Infrared and Raman spectra

# INTRODUCTION

The layered compound $Na_xCoO_2$ represents interesting physics due to the interplay of electronic correlations with lattice degrees of freedom [1]. Its electronic phase diagram is exceptionally rich and includes superconductivity, charge as well as magnetic order, and a regime with very large thermopower. One control parameter is given by the Na content x, which changes both the thickness of the $CoO_2$ layers and the occupation of the Co bands. Going from the extremes of nominal $CoO_2$ (x = 0) to $NaCoO_2$ (x = 1) the Co valence changes formally from $Co^{4+}$ ($3d^5$) with spin $S$=1/2 to $Co^{3+}$ ($3d^6$) with $S$=0. With respect to the Co orbitals this corresponds to $t_{2g}^5$ and $t_{2g}^6$ configurations. The $e_g$ orbitals of Co are at higher energies. According to theoretical work [2], many of the anomalous properties of $Na_xCoO_2$ can be understood in a local picture, by considering the effect of hole doping (x < 1) on the $Co^{4+}$ site on the lattice symmetry and electronic structure of adjacent $Co^{3+}$ sites. The presence of such a hole shifts the $e_g$ orbitals energetically closer to the $t_{2g}$ states, enhances interatomic hopping, and stabilizes the intermediate-spin (S=1) state on the $Co^{3+}$ ions adjacent to $Co^{4+}$ ions. As a result, these $Co^{3+}$ ions couple antiferromagnetically (AF) in a sea of ferromagnetic (FM) correlations. Data from infrared and photoemission spectroscopies provide evidence for such spin polarons [2]. Bandwith renormalization effects leading to sharp spin polaron bands and related effects in transport have also been observed in Co-, Mn- and Ru-based perovskites with three dimensional (3D) electronic structure, the layered compound $(La,Sr)_2MnO_7$, as well as other FM systems in proximity to a metal-insulator transition, e.g. $(Eu,La)B_6$ and EuO [3]. We will discuss the implications of the spin polaron dynamics in $Na_xCoO_2$ below and compare it with these systems.

The interplay of electronic correlations in $Na_xCoO_2$ with the lattice is demonstrated most clearly at certain compositions with commensurable Na vacancy ordering perpendicular to the $CoO_2$ layers. Commensurability effects have been reported for x = 1/3, 1/2, 0.71, and 0.84 [4,5]. From a structural point of view, the superlattice formation is restricted to certain Na vacancy levels with di-, tri- or quadri-vacancy ordering and originates in local strain induced by the Na vacancies. Therefore the doping process is not smooth, but evolves with step-like variations.

For x=0.71 the Na vacancy order stabilizes an unusual electronic state that has been termed *polaronic* or *Curie-Weiss metal*. Here, the AF ordering observed at higher x [6] is suppressed ($T_N \rightarrow 0$), despite a large Curie-Weiss contribution to the magnetic susceptibility. Transport experiments show a T-linear resistivity, $\rho$, a power law in a magnetic field, $\rho(H) \sim H^n$ with 1 < n < 1.6, and a logarithmic divergence of the specific heat at low temperatures. These anomalous properties can be understood as a consequence of a Fermi surface reconstruction leading to the formation of small pockets, and a localization of some of the holes in ordered Na vacancy clusters [7]. Localized and itinerant electrons thus coexist in a manner resembling some 4f and 5f electron compounds [8]. In this situation, geometrical frustration of the localized spins [9] may preclude magnetic order, possibly giving rise to a novel spin liquid state. Interestingly, the polaronic metal with x=0.71 has the same Co valence as the hydrated composition that shows superconductivity [10].

We have performed Raman scattering experiments on phononic and electronic excitations in $Na_xCoO_2$ to test its critical behaviour and the correlation between electronic and lattice degrees of freedom. We first compare a larger set of data on single crystals and thin films to study the dependence of the out-of-plane phonon frequency on Na composition. Then we focus to two compositions, x=0.71 and x=0.84, that share Na vacancy ordering but differ with respect to the supercell size, charge localization, and polaron dynamics. From the observation of electronic Raman scattering a slowing down of the electronic fluctuations and enhanced polarizability with decreasing temperature is deduced. This indicates an unconventional electronic state at low temperatures. Moderate anomalies in frequency and intensity of an out-of-plane phonon point to the importance of out-of-plane coherence for this state.

**EXPERIMENTAL DETAILS**

Raman scattering measurements were performed in quasi-backscattering geometry using $\lambda$ = 514.5 nm $Ar^+$ and $\lambda$ = 532.1 nm solid state lasers. The laser power of P < 10 mW was focused to a d=0.1 mm diameter spot on freshly cleaved (single crystal) or as-prepared (thin film) surfaces. Spectra of the scattered radiation were collected via a triple spectrometer (Dilor XY) and recorded by a nitrogen-cooled CCD detector (Horiba Jobin-Yvon, Spectrum One) with a spectral resolution of $\Delta\omega \sim 0.5$ cm$^{-1}$.

Single crystals and thin films of $Na_xCoO_2$ have been prepared by the optical floating-zone growth method under controlled oxygen atmosphere [4] and by laser ablation, respectively [11,12]. Further experimental characterization has been performed using X-Ray diffraction, transport, and SQUID magnetometry [4,5]. Using electrochemical techniques as a soft chemical reaction, the Na content of selected single crystals has been tuned to reach equilibrium states at x=0.71 and 0.84. In Raman scattering the electrochemical equilibration leads to very sharp phonon modes of the respective single crystals. Previously published Raman work has not considered Na vacancy ordering in detail [13,14].

# RESULTS AND DISCUSSION

The Raman-active oxygen phonon modes of $Na_xCoO_2$ comprise several low frequency ($\omega \sim$ 450 cm$^{-1}$), in-plane modes of $E_g$ symmetry and a higher frequency ($\omega \sim$ 580 cm$^{-1}$), out-of-plane mode of $A_{1g}$ symmetry. The latter mode is of special interest as it couples strongly to electronic correlations on the Co site with momenta close to the center of the Brillouin zone [15]. The low energy, in-plane modes can be used to probe ordering phenomena in the $CoO_2$ planes. Co ion charge ordering leads to a multiplication of low-energy modes as observed, e.g. for x=0.5 [16]. In the presently studied Na-vacancy ordered compositions, two in-plane modes are observed in the frequency range $\Delta\omega \approx 440\text{-}490$ cm$^{-1}$.

The frequency of the out-of-plane mode, given in Fig. 1, shows a continuous variation from 570 to 590 cm$^{-1}$ with Na composition x. This variation is in general agreement with earlier studies [14] and attributed to changes of the Co correlation energy with doping [15]. Data on samples prepared by thin film technology (open squares) are added for completeness. They agree well with single crystal data. Moderate deviations from the linear dependence of the frequency on composition are observed with Na vacancy ordering at x=0.71 and x=0.84 (red circles). We therefore conclude that the Coulomb correlation on Co does not depend strongly on vacancy ordering [4,5].

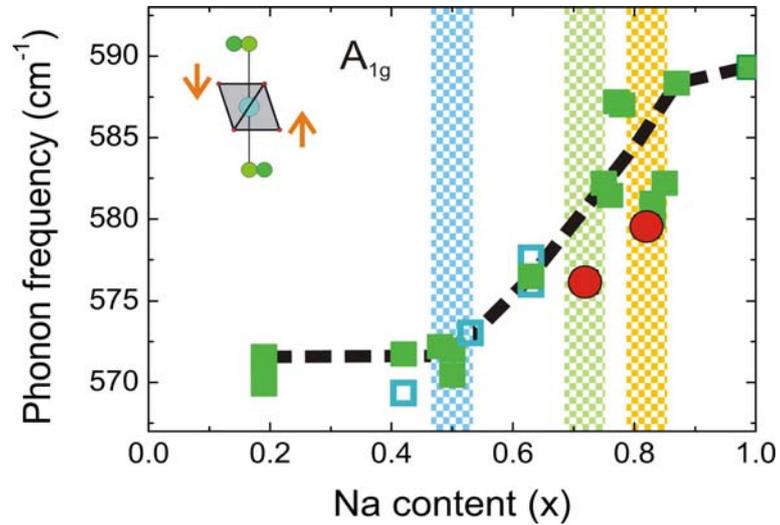

Figure 1: Frequency of the out-of-plane, $A_{1g}$ phonon mode with data from nonequilibrated single crystals (filled squares) and thin films (open squares) and equilibrated, Na ordered single crystals (red circles) at T=10K. The dashed line is a guide to the eye. The dashed bars denote compositions with Na vacancy or Co charge ordering. The inset gives the oxygen ion displacement of the out-of-plane mode with the Co octahedron (blue) and two adjacent Na sites (green).

In the following we will focus on the frequency range below the phonon scattering as shown by Bose corrected Raman data, Im $\chi(\omega)$, in Fig. 2. Low energy Raman scattering is observed with an enormous enhancement of intensity toward low temperatures. This effect decisively depends on composition and vacancy ordering. In the sample with x=0.84 we observe a much smaller intensity and a less pronounced temperature dependence (inset of Fig. 2). This composition also shows a different low energy dynamics in transport and thermodynamic experiments compared to x=0.71 data [5]. The inset of Fig. 2 shows the as-measured, non-Bose-corrected data. This proves that the temperature dependence is by no means a consequence of the Bose factor alone. In samples with Na disorder (x > 0.73) anomalous low-energy scattering has also been observed. Here, however, a plateau of scattering is observed that extends to higher energies. For x≤1 even a maximum at $\omega_{max}$ = 58 cm$^{-1}$ is resolved. This scattering is essentially temperature independent with respect to energy and intensity and has been attributed to collision dominated, electronic Raman scattering [13]. The plateau observed in the disordered samples does not appear to be present in vacancy ordered samples.

In general low energy electronic scattering is observed in metals or doped semiconductors only if dominant charge or spin scattering exist. The collision dominated regime is characterized by a scattering rate $\Gamma$ that is larger than the product of the Fermi velocity and the scattered momentum, $\Gamma > v_F \cdot q$, and the respective scattering intensity is enhanced in proximity to a metal insulator transition [17]. In layered cobaltates the scattering has been attributed to spin-state polarons or other defect states [13]. The maximum position of Bose-corrected data is close to the scattering rate of the process, $\omega_{max} \approx \Gamma$. In the parameter range were the Bose factor is not relevant, the lineshape of the scattering profile is directly related to $\Gamma$. Therefore $\Gamma$ can also be determined by a fit to the high energy shoulder of the data. This is useful if a reduced low energy resolution or elastic scattering do not allow direct detection of the maximum.

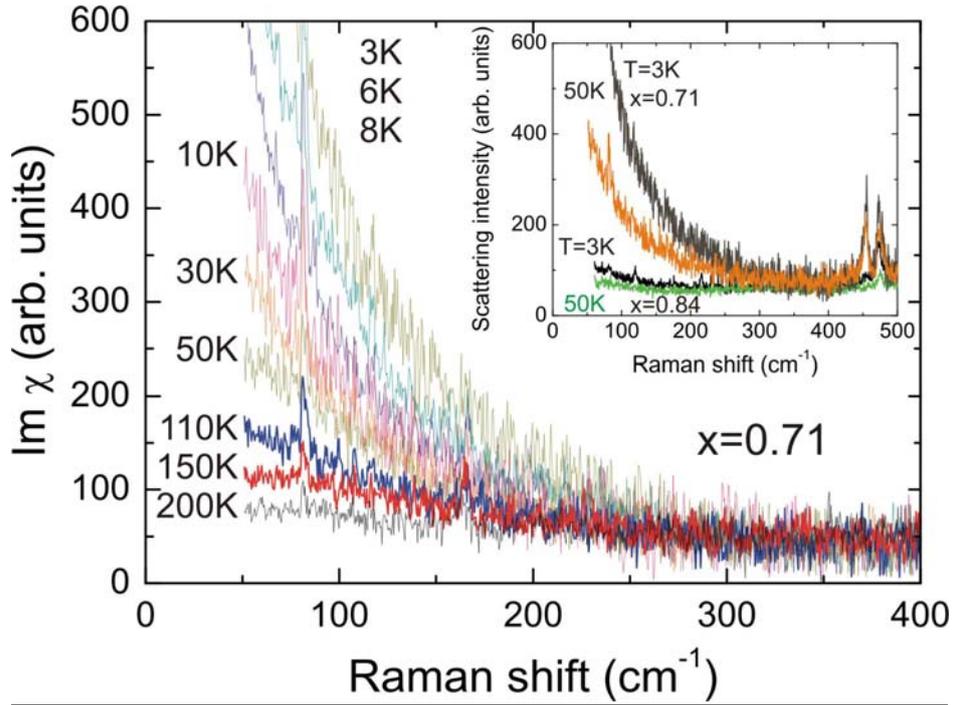

Figure 2: Bose corrected Raman spectra, Im $\chi(\omega)$, for x=0.71 show a low energy scattering intensity that diverges in intensity with decreasing temperatures. The inset compares the Raman scattering intensity at T=3 and 50K for x=0.71 with x=0.84, respectively. The data evidences a kind of critical behavior at low temperatures for x=0.71.

We have performed extensive modeling of our experimental data using Bose-corrected Lorentzian profiles with either $\omega_{max} = 0$ or $\omega_{max} \sim \Gamma \neq 0$. An example for the latter fit is shown in Fig. 3 with data at T = 110K. The Bose-corrected Lorentzian with $\omega_{max} \neq 0$ provides an excellent representation of the data, although the available frequency window does not allow us to resolve the peak position directly. $\Gamma$ determined from the high-energy shoulder is strongly temperature dependent. It shows a fast drop and a power law for T < T* ~ 20 K, $\Gamma(T) \approx A \cdot T^{\alpha}$, with $\alpha \sim 2$, see inset of Fig. 3. A power law means that the fluctuations are not governed by an energy gap. Nevertheless, the meaning of such fitting could be questioned as $\Gamma$ turned out to be smaller than the lower-energy cutoff of the Raman data. Therefore we enhanced the optical resolution of the spectrometer by closing the slit width of the instrument to ½ and ¼ of the nominal value (d=100μm). The low-energy cutoff of the spectrum is now 22 cm$^{-1}$ instead of the previous 50 cm$^{-1}$. The scattering rates determined from these high-resolution data do not differ by more than 3-7 % from those displayed in Fig. 3, depending on temperature. This means that the limited energy resolution has only a small effect on the determination of $\Gamma$ at low temperatures. For higher temperatures we observe a saturation to a value close to the value of Na disordered samples, $\Gamma(T > 200\ K) \approx 60\ cm^{-1}$. This points to a domination of defect-induced electronic scattering processes in this temperature range, as given for Na vacancy disordered samples. Quasi-elastic Lorentzians with fixed $\omega_{max} = 0$ give an equally good description of the data at low temperatures. The derived linewidth $\Gamma_{Lorentz}$ is shown in the inset of Fig. 3. There is also a strong renormalization observed, albeit somewhat less pronounced than in the alternative analysis scheme presented above. Summarizing our evaluation, the low-energy scattering shows a strong renormalization of the linewidth, and therefore of the scattering rate. The temperature dependent linewidth extracted from the data is only weakly dependent on the model used for analysis, and on the restricted frequency window of the experiment.

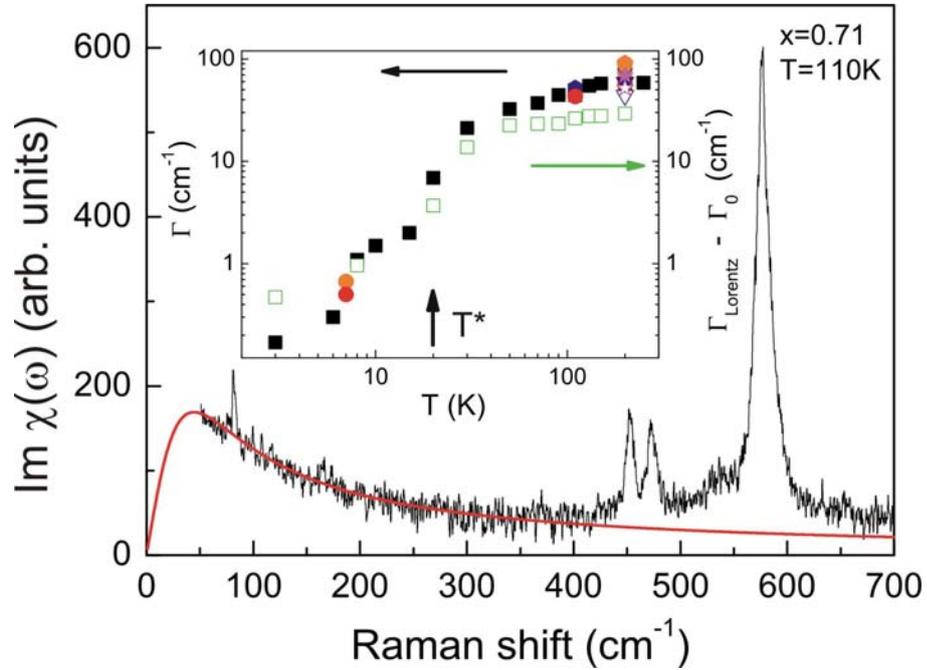

Figure 3: Bose corrected Raman spectrum for Na vacancy ordered x=0.71 at T=110K together with the result of a fit (red line) using a model of collision dominated scattering. The inset shows the scattering rate as function of temperature on a log – log plot. Data with full symbols correspond to a model of collision dominated scattering. Colored symbols show data with higher resolution and a lower energy onset down to 20 cm$^{-1}$. Open symbols give the linewidth $\Gamma_{Lorentz} - \Gamma_0$ with the offset $\Gamma_0 = 80\ cm^{-1}$ of a Lorentzian with $\omega_{max}=0\ cm^{-1}$ [18].

The intensity of the low energy scattering exhibits a similarly large variation at low temperatures, see Fig. 4. A change in slope is visible at T*=20K. The inset of Fig. 4 gives evidence that the low temperature divergence is close to logarithmic. The intensity is determined from a fit using a Lorentzian and a frequency window from 50-400 cm$^{-1}$. This intensity does not depend on the fitting procedure. The observed intensity variation with temperature is also not related to a trivial change of the scattering volume, e.g. via the optical penetration depth or optical constants. This is evident from the invariant intensity of the in-plane phonons that is related to the same geometrical factors, see Fig. 5. We also disregard a collective structural change or Co charge ordering in the low temperature regime as these modes do not change in energy or in number. Therefore we attribute the intensity gain of the collision dominated scattering to an enhanced electronic polarizability of the low temperature state.

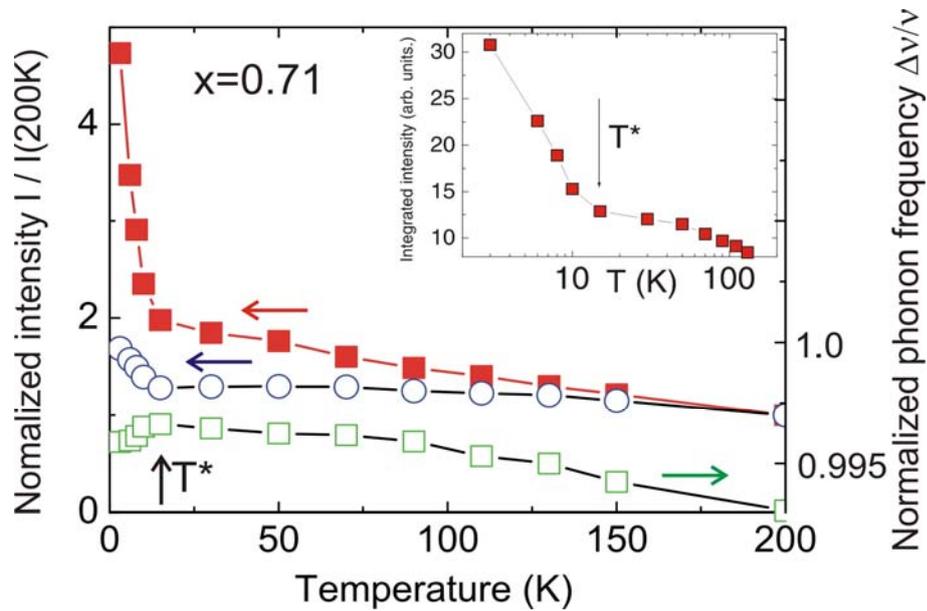

Figure 4: Normalized scattering intensity of the low energy scattering integrated over the energy window 50-400 cm$^{-1}$ (full squares) and the out-of-plane $A_{1g}$ phonon (open circles) scattering together with the $A_{1g}$ phonon frequency shift (open squares, right axis). The inset shows the low energy scattering intensity data on a logarithmic temperature scale. The latter dependence shows similarities to the specific heat attributed a quantum critical point [7].

The out-of-plane phonon mode shows a gain in intensity and the evolution of two sidebands with decreasing temperature. This effect is smaller than the previously discussed intensity increase of the low-energy electronic scattering. In Fig. 4 the data are compared using open circles with filled squares, respectively. It is interesting to note that neither of these effects saturate at the lowest temperatures. The frequency (open squares) of the out-of-plane phonon mode shows a small softening for T< T*. We attribute the effects of the out-of-plane phonon and side band evolution to a crossover to a state with higher lattice coherence along the c axis direction. This takes into account that the vacancy ordering in $Na_{0.71}CoO_2$ involves staging of defect clusters in this direction [5]. Comparing the magnitude of the effects, we conclude that structural degrees of freedom are coupled to the low energy electronic fluctuations, but that are not their main driving force. We note that spin polarons discussed in Refs. 2 are indeed expected to couple to the crystal lattice, because of the different spatial extent of the Co spin states. Further work is required to assess whether this effect can explain the specific phonon anomalies presented here.

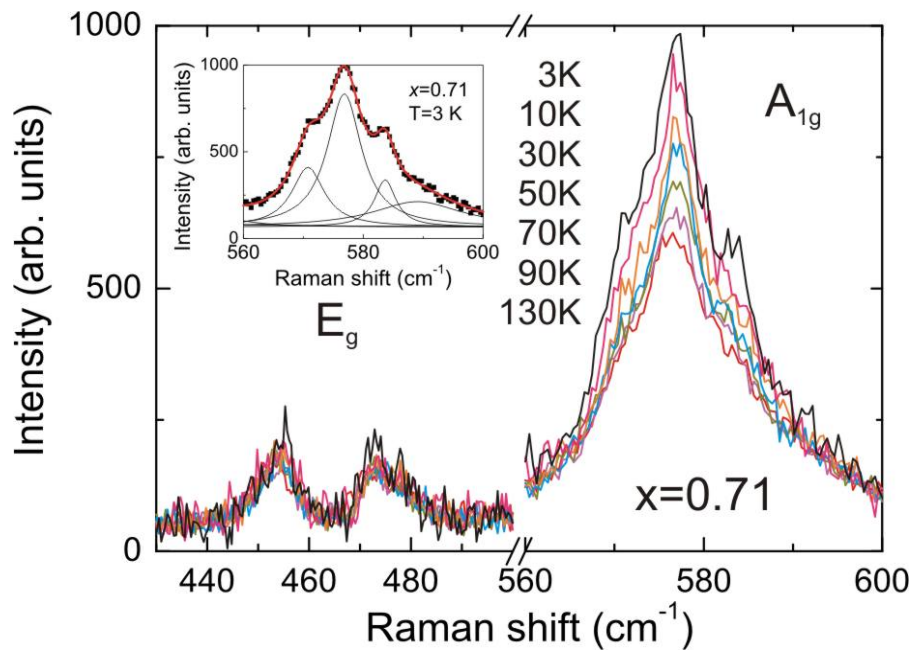

Figure 5: a) Temperature evolution of the $E_g$ in-plane and $A_{1g}$ out-of-plane modes with a continuous intensity increase of the latter and the development of sidebands. In the inset the out-of-plane mode is enlarged together with fits to the main and the side bands.

Our Raman scattering experiments exhibit three major features:

I) collision dominated scattering exists with a scattering rate $\Gamma(T)$ that shows a strong decrease toward low temperatures. The high temperature limit of $\Gamma(T)$ corresponds to the value of vacancy disordered compounds.

II) The low energy scattering intensity, I(T), increases strongly with decreasing temperature, without any indication of saturation.

III) Weaker anomalies are observed in the frequency and intensity of the out-of-plane phonon at T*.

The crossover temperature T* with characteristic changes in all data of Fig. 4 coincides with the ordering temperature, $T_N$ = 18-22 K, of A-type AF order (FM in-plane correlations) observed for disordered samples with x>0.75. [6] For the vacancy ordered phase, however, long range magnetic ordering has been ruled out based on the continuously diverging magnetic susceptibility [5] and power law in the specific heat [7]. Nevertheless, samples with larger x and smaller hole content may show an intrinsically inhomogeneous magnetic state with a spin glass-like freezing of the Co spins at $T_f$=22 K [19]. Therefore, it seems likely that T*=20K for x=0.71 implies a weak change of the spin-spin correlation length. The microscopic origin of this effect remains unclear.

The large reduction of $\Gamma$ and increase of I(T) appear to be directly related. Indeed collision dominated processes with varying peak position and scattering intensity have previously been reported for FM spin polaron systems, as $Eu_{1-x}Gd_xO$ and $Eu_{1-x}La_xB_6$ [17]. Here, maxima exist with scattering rates varying in the range of $\Gamma$=20 – 45 cm$^{-1}$ and 5 – 8.5 cm$^{-1}$, respectively, and are attributed to a metal-insulator transition with enhanced spin-polaron dynamics [17]. In these systems a scaling of the product of temperature and magnetic susceptibility with $\Gamma$(T) has been observed which is due to the dominance of spin fluctuations in the scattering processes with a mean free path l = $\chi \cdot T$. In the regime of increasing scattering rates also the scattering intensities show a moderate gain.

Comparing these cases with $Na_{0.71}CoO_2$, the similarities are obvious, i.e. in the latter the in-plane correlations are FM and spin-orbital polarons have been discussed. However, from $\chi(T) \approx C/T$ (*Curie-Weiss* metal) a constant, temperature independent scattering rate $\Gamma$ would be expected for $Na_{0.71}CoO_2$. Also the temperature variations are opposite: In $Na_{0.71}CoO_2$ the decreasing $\Gamma$(T) is accompanied by an increasing I(T). The decreasing $\Gamma$(T), on the other side, is consistent with the relatively smaller and T-linear resistivity that marks anomalous electronic correlations [5,7]. These observations could point back to the origin of the local moments, a localization of charge carriers to s=1/2 magnetic moments within vacancy clusters and a concomitant restructuring of the Fermi surface to small pockets. The decreasing scattering rate would then be related to a decreasing density of spin polaron states and the increasing electronic scattering intensity to a rather dynamic and highly polarisable low energy state. The major different between $Na_{0.71}CoO_2$ on the one hand and materials such as $Eu_{1-x}Gd_xO$ on the other hand is that the latter show FM order while the former does not exhibit long-range order, probably as a consequence of frustrated interactions [9]. In conjunction with the observation of A-type antiferromagnetism at higher x, this is consistent with a novel quantum critical point.

For this ground state the exact type of Na vacancy ordering plays a decisive role. The ordered compositions x=0.71 and 0.84 differ considerably with respect to the intensity of the collision dominated scattering and the diverging magnetic susceptibility. This contrast is even more pronounced taking also the response of the Na disordered phases into account. Usually electron-phonon leads to instabilities that suppress fluctuations, by opening an excitation gap. For the x=0.71 phase this is not the case. Reasons for the ineffectiveness of electron-phonon coupling are probably related to the complex vacancy cluster ordering that may lead to frustrated magneto-structural interactions and thus prevent the formation of long-range order. Furthermore, the out-of-plane phonon couples mainly to momenta at the centre of the Brillouin zone [15]. It is thereby ineffective to trigger a collective lattice instability that would alter the electronic fluctuations.

Finally, we briefly mention a possible relevance of the dynamic low temperature state of $Na_{0.71}CoO_2$ for the superconducting, hydrated compositions. Hydration of $Na_{0.3}CoO_2$ 1.4 $H_2O$ leads to the intercalation of additional layers separating and decoupling the $CoO_2$ planes. This composition has a Co valence that is very close to the present one in $Na_{0.71}CoO_2$ which implies that the enhanced fluctuations are of relevance for superconductivity [10]. For this composition anomalous electronic scattering processes have also been inferred from transport experiments [20]. On the other side electronic Raman scattering for this composition shows a smaller intensity at low temperatures and a gap formation. This gives evidence that Raman scattering in this case is dominated by a different scattering mechanism and is not dominated by a collision dominated response.

Summarizing, using Raman scattering we give evidence for a regime of pronounced fluctuations for temperatures below T*=20K in the Na vacancy ordered cobaltate $Na_{0.71}CoO_2$. Experimentally, this is based on the observation of a strongly reduced electronic scattering rate and a diverging scattering intensity. The latter corresponds to enlarged electronic polarizability. $Na_{0.71}CoO_2$ differs from other compositions by a suppressed long range magnetic order, anomalous transport and thermodynamic properties and a restructured Fermi surface with a partial localization of charge carriers at vacancy clusters. The observed electronic Raman scattering is interpreted as collision dominated scattering due to ferromagnetically correlated spin-polarons. Ferromagnetic in-plane correlations are also the basis of the observed long range order at higher x. Scenarios of a proximity to a ferromagnetic quantum critical point have been discussed in the literature. Finally, we point out that the phase with anomalous fluctuations can be shifted to a different Na composition (x=2/3) inducing oxygen defects while keeping the Co valence constant [21]. This is consistent with the dominance of electronic correlations over structural effects related to the Na vacancy ordering.

Theoretical studies of the stability of Na vacancy ordered phases [22] as well as on the combination of disorder and correlations on the thermopower [23] have come to similar conclusions.

**ACKNOWLEDGMENTS**

We acknowledge important discussion with P. A. Lee, S. L. Cooper, and D. Wulferding. This work was supported by DFG.